\documentclass[amsmath,amssymb,superscriptaddress,longbibliography,twocolumn]{revtex4-1}
\usepackage{graphicx}
\usepackage{varioref}
\usepackage{xr-hyper}
\usepackage{xcolor}
\usepackage{nicefrac}
\usepackage{xfrac}
\usepackage{hyperref}
\hypersetup{colorlinks,linkcolor=teal,urlcolor=teal,citecolor=orange}
\usepackage{ulem}
\usepackage{siunitx}

\usepackage{subfiles}

\begin{document}

\title{Discovery of Giant Unit-Cell Super-Structure in the Infinite-Layer Nickelate PrNiO$_2$}
\author{J.~Oppliger}
\email{jens.oppliger@physik.uzh.ch}
\affiliation{Physik-Institut, Universit\"{a}t Z\"{u}rich, Winterthurerstrasse 190, CH-8057 Z\"{u}rich, Switzerland}

\author{J.~Küspert}
\affiliation{Physik-Institut, Universit\"{a}t Z\"{u}rich, Winterthurerstrasse 190, CH-8057 Z\"{u}rich, Switzerland}

\author{A.-C.~Dippel}
\affiliation{Deutsches Elektronen-Synchrotron DESY, Notkestra{\ss}e 85, 22607 Hamburg, Germany.}

\author{M.~v.~Zimmermann}
\affiliation{Deutsches Elektronen-Synchrotron DESY, Notkestra{\ss}e 85, 22607 Hamburg, Germany.}

\author{O.~Gutowski}
\affiliation{Deutsches Elektronen-Synchrotron DESY, Notkestra{\ss}e 85, 22607 Hamburg, Germany.}

\author{X. Ren}
\affiliation{Beijing National Laboratory for Condensed Matter Physics, Institute of Physics, Chinese Academy of Sciences, Beijing 100190, China}
\affiliation{School of Physical Sciences, University of Chinese Academy of Sciences, Beijing 100049, China}

\author{X.~J.~Zhou}
\affiliation{Beijing National Laboratory for Condensed Matter Physics, Institute of Physics, Chinese Academy of Sciences, Beijing 100190, China}

\author{Z.~Zhu}
\affiliation{Beijing National Laboratory for Condensed Matter Physics, Institute of Physics, Chinese Academy of Sciences, Beijing 100190, China}

\author{R.~Frison}
\affiliation{Physik-Institut, Universit\"{a}t Z\"{u}rich, Winterthurerstrasse 190, CH-8057 Z\"{u}rich, Switzerland}

\author{Q.~Wang}
\affiliation{Physik-Institut, Universit\"{a}t Z\"{u}rich, Winterthurerstrasse 190, CH-8057 Z\"{u}rich, Switzerland}
\affiliation{Department of Physics, The Chinese
University of Hong Kong, Shatin, Hong Kong, China}

\author{L.~Martinelli}
\affiliation{Physik-Institut, Universit\"{a}t Z\"{u}rich, Winterthurerstrasse 190, CH-8057 Z\"{u}rich, Switzerland}

\author{I.~Bia\l{}o}
\affiliation{Physik-Institut, Universit\"{a}t Z\"{u}rich, Winterthurerstrasse 
190, CH-8057 Z\"{u}rich, Switzerland}

\author{J.~Chang}
\email{johan.chang@physik.uzh.ch}
\affiliation{Physik-Institut, Universit\"{a}t Z\"{u}rich, Winterthurerstrasse 190, CH-8057 Z\"{u}rich, Switzerland}


\maketitle

\textbf{Spectacular quantum phenomena such as superconductivity often emerge in flat-band systems where Coulomb interactions overpower electron kinetics. Engineering strategies for flat-band physics is therefore of great importance. 
Here, using high-energy 
grazing-incidence x-ray diffraction, 
we demonstrate how \textit{in-situ} temperature annealing of the infinite-layer nickelate PrNiO$_2$ 
induces a giant superlattice structure.  The annealing effect has a maximum 
well above room temperature. By covering a large scattering volume, we show a rare 
period-six in-plane (bi-axial) symmetry and a 
period-four symmetry in the out-of-plane direction. 
This giant unit-cell superstructure 
likely stems from ordering of diffusive oxygen.
The stability of this superlattice structure suggests a connection to an energetically favorable electronic state of matter. As such, our study provides a new pathway -- different from Moiré structures -- to ultra-small Brillouin zone electronics.} 
\\[4mm]

\begin{figure*}
\includegraphics[width=0.995\textwidth]{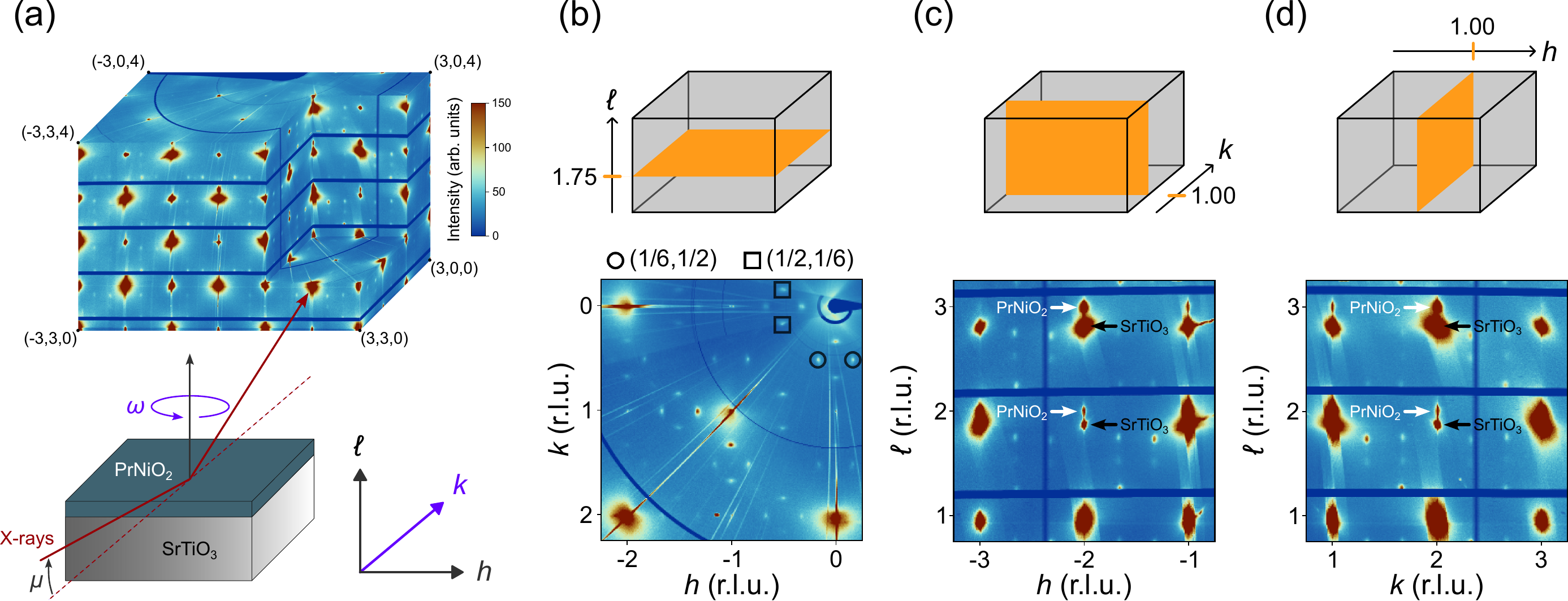}
\caption{\textbf{High-energy grazing-incidence x-ray diffraction on a PrNiO$_2$ thin film.} (a) Schematic illustration of diffraction geometry. A three-dimensional scattering volume is recorded by high-energy grazing-incidence (angle $\mu$) x-rays, diffracted on a horizontal film, which is rotated around its vertical axis ($\omega$). (b-d) Two-dimensional cuts displayed schematically (top) together with intensity maps of 
respectively the $(h,k,1.75)$, $(h,1,\ell)$ and $(1,k,\ell)$ scattering planes measured at 386~K (bottom). Diffracted intensities are visualized using a linear false color scale. Fundamental Bragg reflections of the PrNiO$_2$ film and the SrTiO$_3$ (STO) substrate are indicated by arrows in (c) and (d). Principle superlattice reflections are highlighted with square and circular symbols in (b).}
\label{fig:figure_3D}
\end{figure*}

Applications of transition metal oxides span from dental restoration to high-tech semiconductor devices~\cite{ShiAM2021}. At the same time, 
oxide materials host some of the most enigmatic phases of quantum matter. For example, high-temperature superconductivity in the cuprates (copper-oxides) is still an active field of research~\cite{Keimer2015}. A longstanding challenge is to -- by design -- realize 
cuprate-physics in other materials~\cite{Norman2016}. Low-valence nickelates have been a prime candidate for this task. The discovery of superconductivity in doped La$_{1-x}$Sr$_x$NiO$_2$ therefore sparked immediate excitement~\cite{LiNature2019,LiPRL2020,OsadaPRM2020,ZengPNAS2022,ZengPRL2020}. Much of the following experimental work  has been discussed with cuprate physics as reference~\cite{NormanPRX2020,Mitchell2021, goodge2021}. Experimental studies and calculations agree on a dominant $3d^{9-\delta}$ ground state, but highlighted important differences with respect to cuprates, including a more prominent Mott-Hubbard gap and an active role of rare-earth bands at the Fermi level \cite{hepting2020, kitatani2020a, NormanPRX2020, goodge2021}.

Similarities were strengthened by the discovery of dispersive magnon excitations, revealing 
strong antiferromagnetic exchange~\cite{LuScience2021, KriegerPRL2022}.
A crucial characteristics of cuprates is the presence of two-dimensional charge order in the superconducting planes. Such modulation, ubiquitous both in hole~\cite{tranquada, chang12, ghiringhelli2012long}  and electron-doped cuprates~\cite{dasilvanetoe2014ubiquitous}, seems to be a rather fundamental property of the two-dimensional Hubbard model~\cite{jiang2019superconductivity, marinoStripesExtendedTt2022}. Therefore, great experimental effort has been put in the search of a similar broken symmetry in nickelates.
Recently, the presence of a charge modulation along Ni-O bonds, 
was discovered in La-,  Nd- and Pr-based nickelates 
by resonant 
x-ray scattering~\cite{RossiNatPhys2022, KriegerPRL2022,TamNAtMAt2022,ren_two_2024_2,rossi_universal_2024}. However, 
the interpretation of these results in terms of charge order is surrounded by controversy~\cite{pelliciari2023comment,parzyck2024absence}.
Unlike in cuprates, the order lacks a clear temperature dependence~\cite{KriegerPRL2022, TamNAtMAt2022}. Moreover, its dependence on sample preparation~\cite{TamNAtMAt2022} and the unclear role of an epitaxial capping layer~\cite{KriegerPRL2022} question its universality in the family of nickelates.
Other proposals to explain the observed modulation include the formation of hydrogen chains~\cite{liang2023topo} or superstructure of re-intercalated oxygen atoms~\cite{raji2023charge}. 


Here we present a high-energy, 
grazing-incidence x-ray diffraction study of PrNiO$_2$ with crystalline and amorphous SrTiO$_3$ (STO) capping layer. 
In contrast to resonant diffraction, this technique covers a large scattering volume across many Brillouin zones. Our main finding is a stable, giant unit cell emerging 
upon \textit{in-situ} 
thermal heating 
above ambient temperature. In the NiO$_2$ plane, a rare period-six 
translational symmetry occurs with a 
period-four stacking order in the out-of-plane direction. 
This giant unit-cell superstructure remains stable over 
a large temperature range and emerges irrespectively of crystalline or amorphous capping. As such, it represents a fundamentally novel structure -- most likely originating from ordering of diffusive oxygen. Quenching this structure to low-temperature 
promises access to new ultra-small Brillouin zone electronics. 


    
\section*{Results}

Our grazing-incidence diffraction geometry is schematically illustrated in the lower part of Fig.~\ref{fig:figure_3D}(a). 
By rotating the sample around the direction perpendicular to the scattering plane,
a three-dimensional scattering volume is collected. 
As exemplified by data on a PrNiO$_2$ thin film grown on an STO substrate with crystalline STO capping layer, this scattering volume $(h,k,\ell)$ covers dozens of Brillouin zones as shown in the upper part of Fig.~\ref{fig:figure_3D}(a). In Fig.~\ref{fig:figure_3D}(b-d), we display two-dimensional slices of the scattering volume. From such slices, fundamental Bragg reflections yield information about lattice parameters and translational symmetry breaking.
As 
common for expitaxially strained growth, the in-plane lattice parameters of PrNiO$_2$ and the STO substrate are 
identical within our experimental resolution (see Table~\ref{tab:tab1} of the methods section).
By contrast, along the out-of-plane $c$-axis direction, lattice parameters of substrate and the PrNiO$_2$ film are clearly different 
-- see Fig.~\ref{fig:figure_3D}(c,d). 

\begin{figure*}
\includegraphics[width=0.995\textwidth]{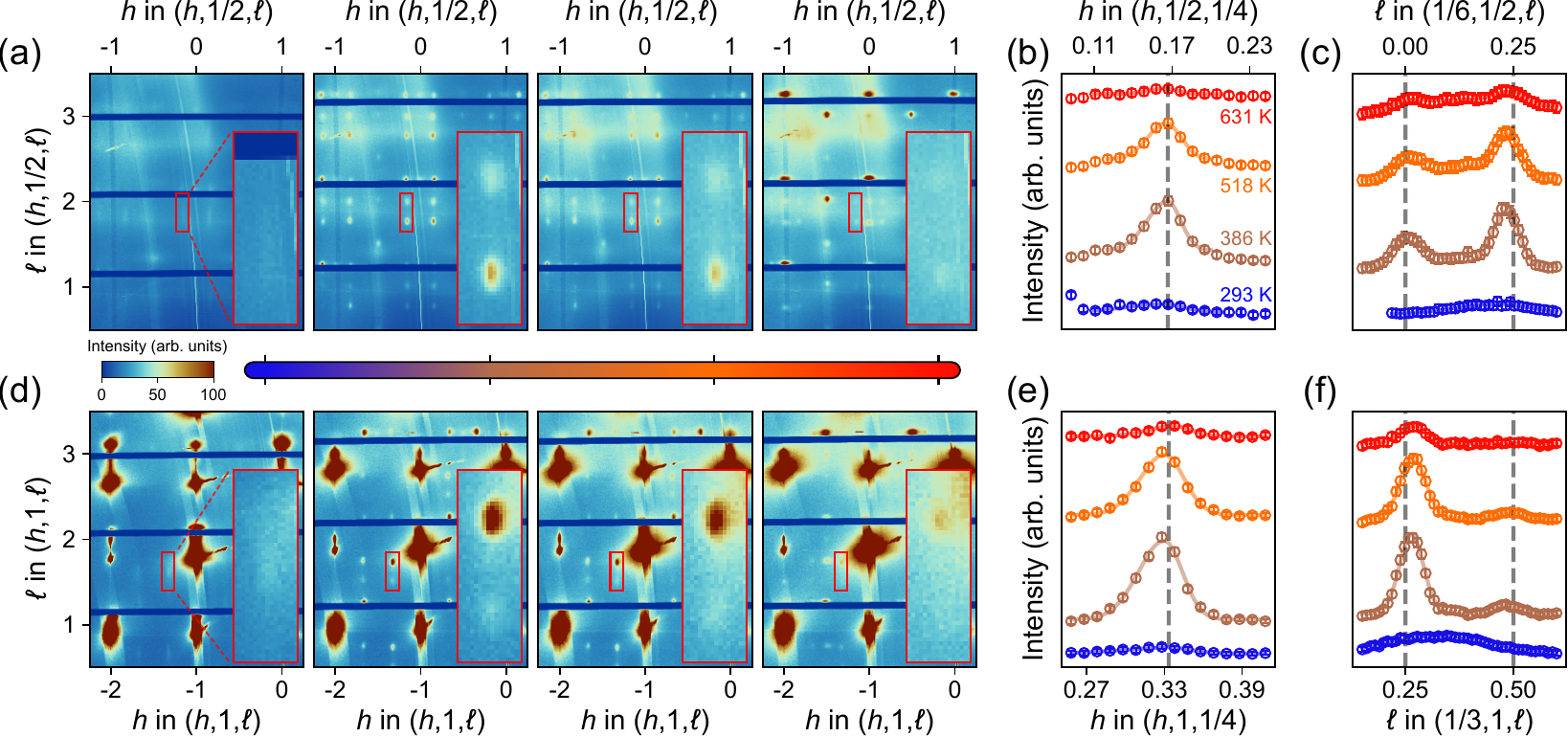}
\caption{\textbf{
Thermal-induced superlattice structure in a PrNiO$_2$ thin film.} (a,d) Diffraction intensities (linear false color scale) in the $(h,1/2,\ell)$ and $(h,1,\ell)$ scattering planes as a function of temperature. The four temperatures are indicated in panel (b). The most intense peaks stem from fundamental Bragg peaks of the STO substrate and the PrNiO$_2$ thin film. Selected superlattice peaks 
are highlighted  by the red rectangular boxes. (b,c) One-dimensional $h$ (in-plane) and $\ell$ (out-of-plane) scans through 
the superlattice reflections in (a) for temperatures as indicated. (e,f) Equivalent $h$ and $\ell$ scans but through 
the superlattice reflections in (d). Solid lines are Gaussian profiled fits  with a sloping background. Error bars reflect counting statistics.} 
\label{fig:figure_2D}
\end{figure*}

We find that fundamental Bragg peaks at $\tau~=~(h_B,k_B,\ell_B)$ -- with $h_B$, $k_B$, and $\ell_B$ being integers -- are virtually 
temperature independent. 
Upon heating above room temperature, we discover the emergence of 
additional 
commensurate reflections. These reflections occur at
$Q_{o} =  \tau q_{o}^{i}$ with

 
%
%
%

\vspace*{-0.4cm}
\begin{equation}
    q_{o}^{a,b} = (\delta_h,\delta_k,0) \quad \textrm{and} \quad 
q_{o}^{c} = (0,0,\delta_\ell)
\end{equation}




and in-plane commensurabilities $\delta_h~\approx~\delta_k~\approx~1/6$ and out-of-plane commensurability $\delta_\ell \approx 1/4$. 
Examples of superlattice peaks at $Q_o~=~(\pm 1/6,1/2,7/4)$ and $Q_o~=~(1/2,\pm 1/6,7/4)$ are highlighted by circles and squares in Fig.~\ref{fig:figure_3D}(b).
Notice that the  $Q_o~=~(1/6,1/6,\ell_B \delta_\ell)$ reflection is either weak or symmetry forbidden. 

In Fig.~\ref{fig:figure_2D}, we 
focus on the $(h,1/2,\ell)$ and $(h,1,\ell)$ scattering planes respectively for a PrNiO$_2$ thin film with crystalline STO capping layer -- see Supplementary Fig.~\ref{fig:figure_S1} for equivalent $(1/2,k,\ell)$ and $(1,k,\ell)$ data. 

Reflections in both scattering planes display the same temperature dependence. Initially, the superlattice peaks emerge and are enhanced upon heating above room temperature. Note that at room temperature, the out-of-plane scan reveals 
a broad peak centred around $\delta_\ell \approx 1/3$ -- see Fig.~\ref{fig:figure_2D}(f). This is also in agreement with 
previous resonant x-ray scattering studies~\cite{TamNAtMAt2022,parzyck2024absence,ren_two_2024_2}.
Upon heating, the out-of-plane commensuration changes to a sharp peak with $\delta_\ell \approx 1/4$ as shown in Fig.~\ref{fig:figure_2D}(c,f). 
This phase with quarter commensuration furthermore displays a much longer out-of-plane correlation length $\xi_c$, indicating an improved stacking order. 
The PrNiO$_2$ film with amorphous capping does not display any ordering at room temperature. For this system, the superlattice structure emerges only at higher temperatures first with $\delta_h \approx \delta_k \approx 1/6$ and a broad peak with $\delta_\ell \approx 1/3$. At even higher temperatures the out-of-plane commensuration -- like in the case of crystalline capping -- manifests as a sharp peak with $\delta_\ell \approx 1/4$. 

\begin{figure*}
\includegraphics[width=0.995\textwidth]{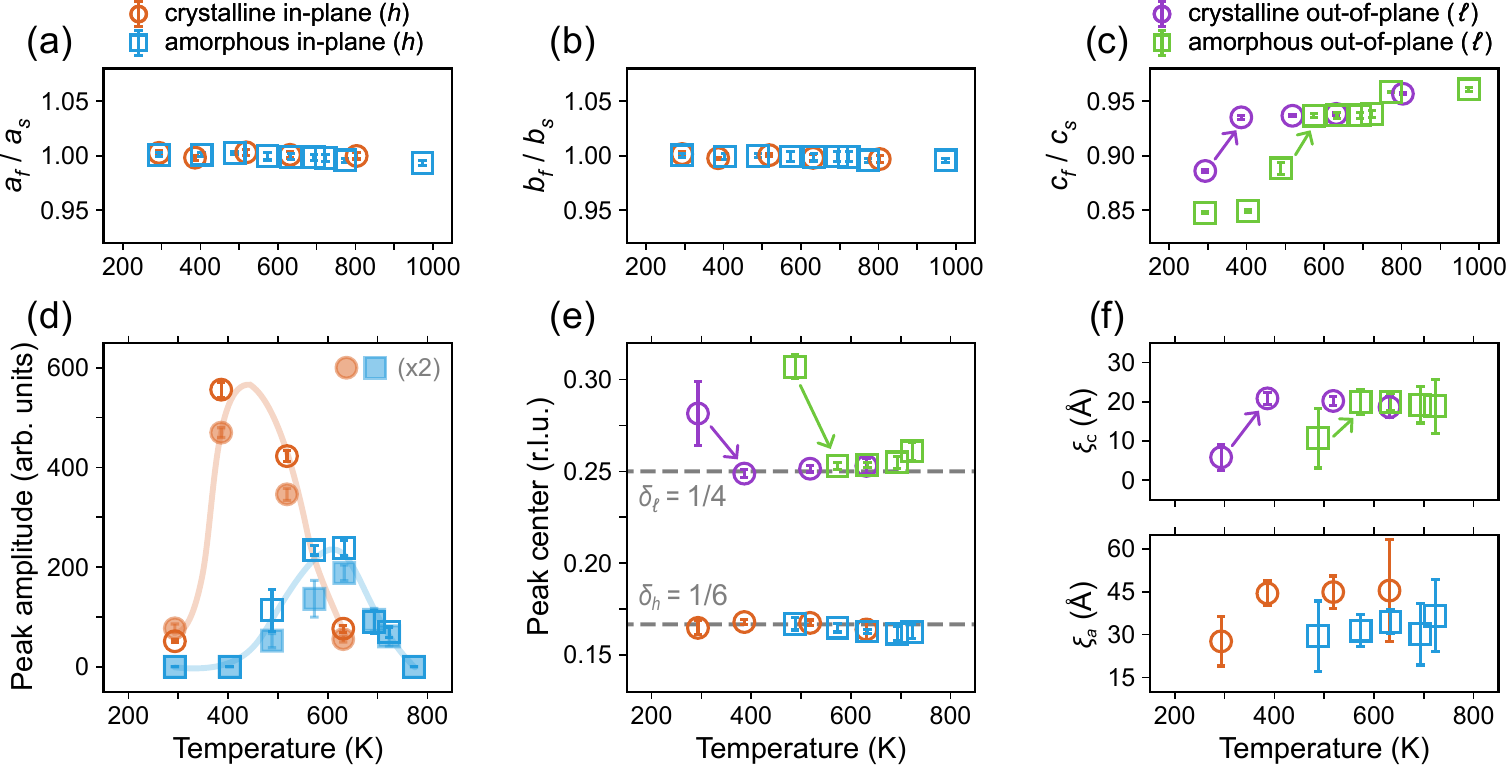}
\caption{\textbf{Temperature and capping layer dependence of the superlattice structure.} (a-c) PrNiO$_2$ thin film lattice constants $x_f$ -- normalized to the STO substrate lattice constants $x_s$, with $x = a,b,c$ for films with crystalline and amorphous capping layers. (d) Peak amplitude versus temperature of selected 
reflections 
$Q_{o}~=~(1/6,1/2,7/4)$ (filled markers)
and $Q_{o}~=~(1/3,1,7/4)$ (empty markers)
for films with crystalline and amorphous STO capping. Amplitude of $Q_{o} = (1/6,1/2,7/4)$ has been scaled by a factor of two. Solid lines are guides to the eye.
(e) In-plane $\delta_h$ and out-of-plane $\delta_\ell$ commensuration plotted versus temperature.
(f) In-plane $\xi_{a}$ and out-of-plane $\xi_{c}$ correlation lengths versus temperature. Data in (e) and (f) is averaged over the two reflections 
separately for crystalline and amorphous capping. Error bars represent one standard deviation obtained from a least squares fitting procedure.}
\label{fig:figure_1D}
\end{figure*}

In Fig.~\ref{fig:figure_1D}, we summarize the temperature dependence of the superlattice peaks for systems with crystalline and amorphous STO capping layer.
Fig.~\ref{fig:figure_1D}(a,b) shows that the PrNiO$_2$ in-plane $(a,b)$ lattice constants 
are essentially temperature independent. By contrast, the $c$-axis lattice constant shows a step-like temperature dependence -- see Fig.~\ref{fig:figure_1D}(c). Roughly at this step (highlighted with arrows), the quartet peaks at $Q_o$ appear.
For temperatures  above 600~K these peaks are suppressed and eventually vanish at temperatures above 800~K as shown in Fig.~\ref{fig:figure_1D}(d). High enough temperatures therefore seem to reverse the topotactic reaction and return the film system to the PrNiO$_3$ cubic perovskite structure. This is confirmed by 
laboratory 
2$\theta$ scans shown in Supplementary Fig.~\ref{fig:figure_S2}.
These trends are observed both for films with crystalline and amorphous capping.
For the sample with amorphous capping, the onset temperature of the quartet peaks is shifted by around 100 K and is less pronounced compared to the sample with crystalline capping. Interestingly, the in-plane $\delta_h \approx \delta_k \approx 1/6$ and out-of-plane $\delta_\ell \approx 1/4$ 
commensurations 
show little to no temperature dependence
-- see Fig.~\ref{fig:figure_1D}(e). Similarly, 
both in-plane and out-of-plane correlation lengths -- shown in Fig.~\ref{fig:figure_1D}(f) -- plateau in the ordered state. 

\section*{Discussion}

\begin{figure}
\includegraphics[width=0.4\textwidth]{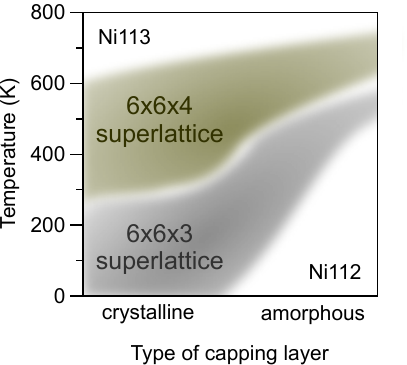}
\caption{\textbf{Schematic "phase" diagram of ANiO$_{3-x}$.} Through a topotactic reduction, ANiO$_3$ (Ni113) films can be oxygen reduced to stochiometric ANiO$_2$ (Ni112) that upon hole doping displays superconductivity. Our study suggests that oxygen diffusion can stabilize a 6x6x4 superstructure. The diffusion process can be accelerated by annealing the thin film. At high enough temperatures, the system will return to the ANiO$_{3-x}$ oxygen stochiometry. }
\label{fig:figure_schematic}
\end{figure}

We summarize our results using a schematic diagram in Fig.~\ref{fig:figure_schematic}. Between the known ANiO$_2$ and ANiO$_3$ crystal structures, there exist -- at least two -- superlattice structures with gigantic unit cells. The superlattices are composed of two independent orderings: A fundamental two-dimensional 
ordering and different stacking patterns. This is reminiscent of two-dimensional charge orderings in the cuprates or dichalcogenides where 
different stacking orders 
frequently occur. In our particular case, we report a fundamental 
in-plane order 
that stacks with a (short-range) period three or a (long-range) period four 
along the $c$-axis. Based on our diffraction experiment, it is not possible to  distinguish checkerboard (biaxial) from twinned stripe order.

Irrespective of exact symmetry breaking, the reflections 
contain valuable information about the nature of the ordering. The 
observed superlattice reflections are intense -- only one or two orders of magnitude weaker than the fundamental Bragg peaks of the thin film (see Fig.~\ref{fig:figure_3D}). This suggests that the symmetry breaking stems from a strong ordering tendency~\cite{JaramilloNature2009}. 
This would be atypical for charge density waves that typically manifest by weak reflections. Yet, both electronic or atomistic driven mechanisms are possible.

The fact that the ordering can be quenched 
(see Supplementary Fig.~\ref{fig:figure_S3}), 
suggests that the observed symmetry breaking goes beyond a standard crystal structure phase transition. 
It is possible that oxygen diffuses from the substrate and/or capping layer 
to the PrNiO$_2$ film or that the topotactic process resulted in a residual apical oxygen occupation.
High temperatures 
will 
enhance oxygen diffusion and 
promote an oxygen annealing process as seen for example in YBa$_2$Cu$_3$O$_{6+x}$~\cite{zimmermann_oxygen-ordering_2003}.
Oxygen diffusion would render our PrNiO$_2$ film off-integer stochiometric by occupying vacant apical oxygen positions. Such a partial apical oxygen occupation is consistent with the observed $c$-axis extension -- see Fig.~\ref{fig:figure_1D}(c).
A single apical oxygen atom per 6x6x4 (original) unit cells generates the observed symmetry breaking. As such, the reported superlattice structure is closely related to the ANiO$_2$ composition. 


We stress that due to the weak form factor, apical (or in-plane) oxygen alone 
can not explain the observed structure factor. However, apical oxygen inclusions may induce Ni and Pr distortion patterns. Due to the large atomic mass, Pr distortions are likely to dominate the structure factor. A structural refinement would be an interesting future extension of this work.



An open 
pressing question is as to why the 
giant 6x6x4 unit cell 
manifests over a 300~K temperature range, 
irrespective of crystalline or amorphous capping. In principle, oxygen diffusion 
would produce an arbitrary oxygen stochiometry. Our observation of a stable 
giant superstructure implies a significant 
down-scaling of the Brillouin zone. This in turn induces massive band folding that often generates flat band physics as for example seen in magic-angle twisted bilayer graphene~\cite{balents_superconductivity_2020}.
As such, it is possible that the superstructure induces an electronic state with favorable energetics.
This hypothesis therefore implies the existence of two fundamentally different ground states of PrNiO$_{2+x}$. 
The fact that spin excitations -- in ANiO$_2$ -- are not observed in combination with this symmetry breaking~\cite{KriegerPRL2022}, supports this rationale. It would thus be of great interest to quench the giant superlattice structure to low temperature for studies of its electronic structure and properties.

\section*{Methods}{\label{methods}}

\textit{Film systems:}
Two different capped films of PrNiO$_2$ on a SrTiO$_3$ (STO) substrate have been studied. Thicknesses of films and cappings are indicated in Table~\ref{tab:tab1} along with lattice parameters. The thin films were grown on a (001)-oriented SrTiO$_3$ substrate by pulsed laser deposition. During growth, the substrate temperature was kept at 600~°C under an oxygen partial pressure of 150 mTorr. After topotactic reduction (390~°C for 2~h), the pervoskite phase is transformed into an infinite-layer phase.
The dimensions of both samples are 5$\times$5$\times$0.5 mm$^3$. Both samples were cleaned with isopropyl and afterwards dried with compressed air. Prior to the measurement the thin film with crystalline capping was exposed to air for over a day. The thin film with amorphous capping was measured immediately after it has been removed from an inert atmosphere.\\ 

\begin{table}[ht!]
\caption{\textbf{Studied film systems.} Lattice parameters and thicknesses of the two capped film systems used for this study. In-plane parameters of the substrate, film and capping are identical within the experimental sensitivity. Given that the capping layer is very thin, it is not possible to identify the $c$-axis lattice parameter and hence this entry is indicated by $\varnothing$. Use of amorphous STO capping yields a $c$-axis lattice parameter comparable to cap-free PrNiO$_2$ films~\cite{osada_superconducting_2020_2,OsadaPRM2020}.}

\vspace{2mm}
\begin{ruledtabular}
\begin{tabular}{ccccc}
Film system & Thickness [nm] & a [\AA] & b [\AA]& c [\AA]\\ 
\hline
Cryst. STO capping & 4.0 & $\vert$ & $\vert$ & $\varnothing$\\
PrNiO$_2$&  7.0 & 3.885 & 3.905 & 3.450\\
STO substrate & $\infty$ & $\vert$ & $\vert$ & 3.895\\
\hline
Amorph. STO capping & 5.0 & $\vert$ & $\vert$ & $\varnothing$\\
PrNiO$_2$&  7.6 & 3.900 & 3.915 & 3.305\\
STO substrate & $\infty$ & $\vert$ & $\vert$ & 3.900
\end{tabular}
\end{ruledtabular}
\label{tab:tab1}	
\end{table}

\textit{Diffraction experiments:} 
High energy x-ray diffraction experiments were carried out at the second experimental hutch (EH2) of the P07 beamline~\cite{Schell2010TheHE,gustafson_high-energy_2014,bertram_1d_2016} at the PETRA III 
storage ring (DESY, Hamburg). 73~keV x-rays with grazing-incidence geometry ($\mu = 0.05^\circ$) and a Detectris Pilatus3 X CdTe 2M detector were used. For each scan, an angular range of 200$^\circ$ ($\omega$ in Fig.~\ref{fig:figure_3D}(a)) has been covered using a total of 2000~frames. Each frame therefore corresponds to an angular range of 0.1$^\circ$. The exposure time per frame was set to 0.05~s. Temperature was controlled by a resistive heating plate and the sample was kept in a helium atmosphere with constant flow rate.\\

\textit{Data analysis:}
Detector images are reconstructed into reciprocal space and shown two-dimensional data slices are integrated over 0.1 reciprocal lattice units (r.l.u.) along the slicing direction. The peaks of the one-dimensional line profiles for the sample with (crystalline) amorphous capping layer are fitted with a (linear) quadratic background and a (split) Gaussian function. Correlation lengths when using a split Gaussian function are obtained from the average of the standard deviations of the Gaussians.\\

\textit{Data availability.}
All experimental data are available upon reasonable request to the corresponding authors. \\

\textit{Acknowledgments:}
J.O., J.K., L.M., and J.C. acknowledge support from the Swiss National Science Foundation (200021\_188564). J.O. acknowledges support from a Candoc grant of the University of Zurich (Grant no. K-72334-06-01). J.K. is further supported by the PhD fellowship from the German Academic Scholarship Foundation. I.B. and L.M. acknowledges support from the Swiss Government Excellence Scholarship. Z.Z. acknowledges the support from the National Natural Science Foundation of China (Grant No.~12074411). Q.W. is supported by the Research Grants Council of Hong Kong (ECS No. 24306223), and the CUHK Direct Grant (No. 4053613). Parts of this research were carried out at beamline P07 at DESY, a member of the Helmholtz Association (HGF). 
The research leading to this result has been supported by the project CALIPSOplus under the Grant Agreement 730872 from the EU Framework Programme for Research and Innovation HORIZON 2020.\\

\textit{Author contributions:} X.R., X.J.Z. and Z.Z. grew the PrNiO$_2$ films. Sample preparation for the x-ray experiments were organized by I.B. and J.K.. J.O., J.K., O.G., A.C.D., M.v.Z., and J.C. carried out the experiment. J.O. carried out the data analysis with assistance from M.v.Z., J.K., I.B., L.M., and J.C. The project was conceived by Q.W. and the manuscript was written by J.O. and J.C. with assistance from all authors. J.O. and J.K. contributed equally.\\ 

\textit{Competing interests:} The authors declare no competing interests.\\

\vspace{2mm}

\bibliography{XRD}

\newpage
\
\newpage

\setcounter{figure}{0}
\renewcommand{\figurename}{FIG.}
\renewcommand{\thefigure}{S\arabic{figure}}



\begin{minipage}[c]{2\columnwidth}
\section*{Supplementary information}
\end{minipage}

\vspace{1cm}

\begin{figure}[ht]
\begin{minipage}[c]{2\columnwidth}
\includegraphics[width=0.995\textwidth]{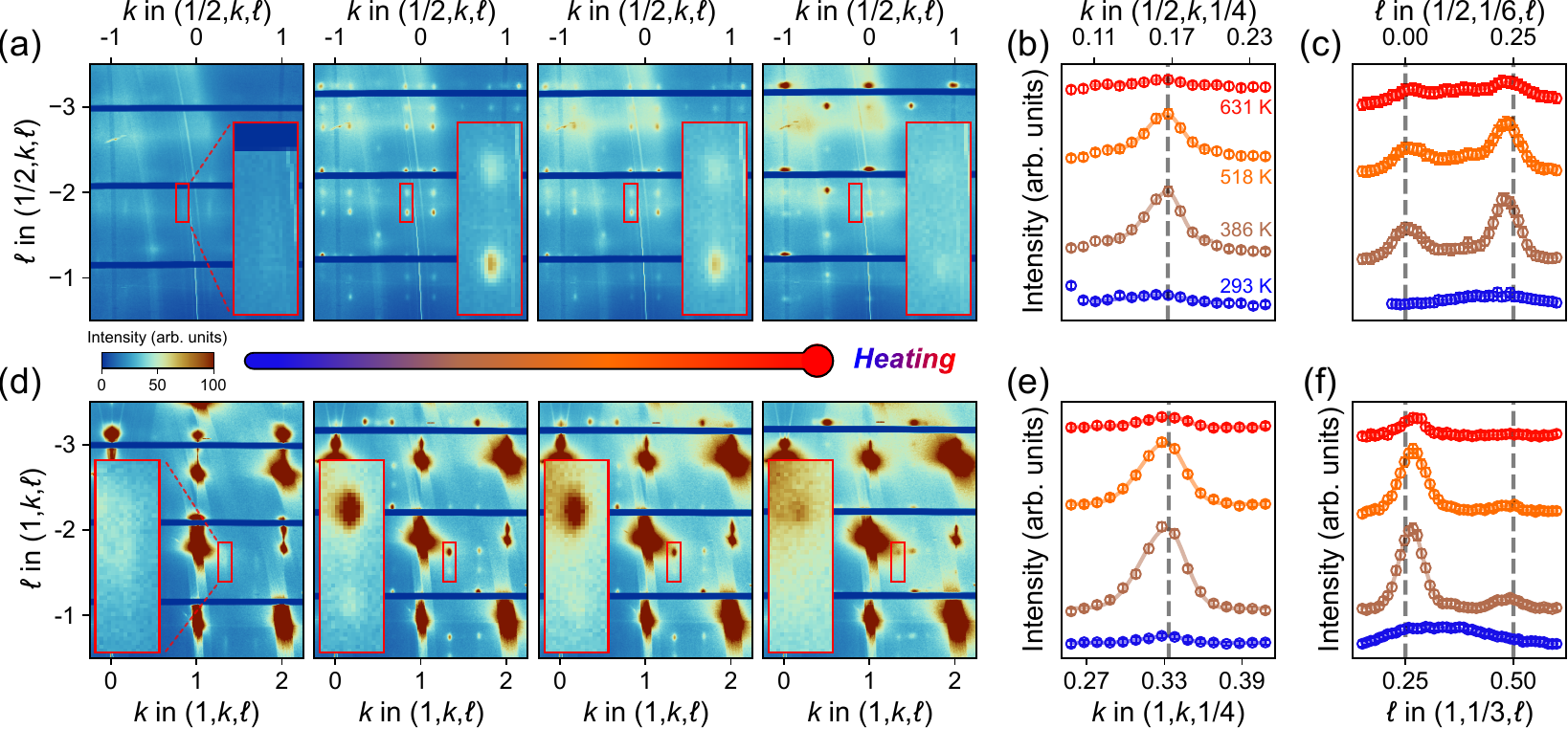}
\caption{\textbf{
Thermal-induced superlattice structure in a PrNiO$_2$ thin film.} (a,d) Diffraction intensities (displayed with a linear false color scale) in the $(1/2,k,\ell)$ and $(1,k,\ell)$ scattering planes as a function of temperature. The most intense peaks stem from fundamental Bragg peaks of the STO substrate and the PrNiO$_2$ thin film. Selected superlattice peaks are highlighted by red rectangular boxes. (b,c,e,f) One-dimensional $k$ (in-plane) and $\ell$ (out-of-plane) scans through the superlattice reflections for temperatures as indicated. Solid lines are Gaussian profiled fits  with a sloping background. Error bars reflect counting statistics.}
\label{fig:figure_S1}
\end{minipage}
\end{figure}

\begin{figure}[ht]
\begin{minipage}[c]{2\columnwidth}
\includegraphics[width=0.6\textwidth]{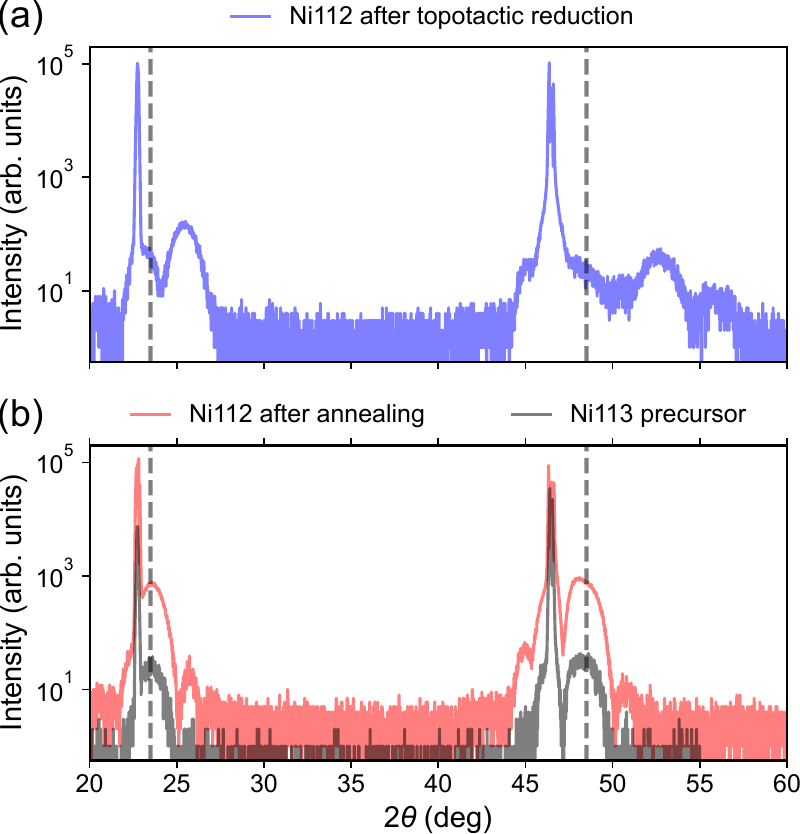}
\caption{\textbf{Room temperature 2$\theta$ scans recorded on PrNiO$_2$ (Ni112) and PrNiO$_3$ thin films with crystalline STO capping layer}. (a) Infinite layer phase Ni112 obtained after topotactic reduction of the perovskite precursor PrNiO$_3$ (Ni113). (b) Comparison of the Ni112 thin film after annealing at high temperatures with the original Ni113 phase. The scans were performed using a SmartLab x-ray diffractometer with a Cu K$\alpha$ source.} 
\label{fig:figure_S2}
\end{minipage}
\end{figure}

\begin{figure}[ht]
\begin{minipage}[c]{2\columnwidth}
\includegraphics[width=1\textwidth]{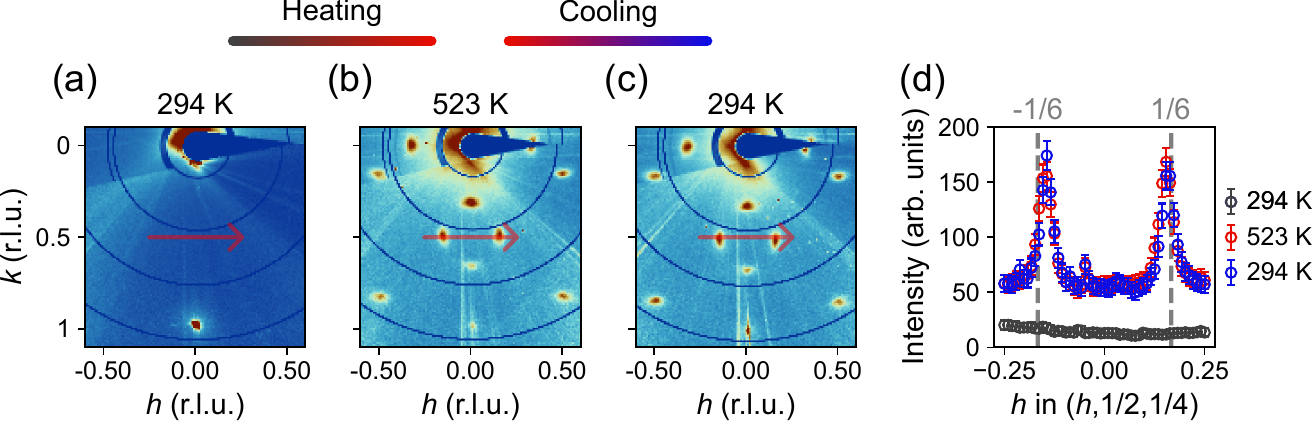}
\caption{\textbf{Temperature quenching of a PrNiO$_2$ thin film with crystalline STO capping layer}. (a-c) Diffracted intensities within the $(h,k,1.75)$ scattering plane for temperatures as indicated. (d) Corresponding $h$-scans through 
$(h,1/2,1/4)$, demonstrating how the annealing-induced symmetry breaking can be quenched. This quenching experiment was carried out on a sample that was kept under vacuum conditions before being introduced into the controlled helium atmosphere in the XRD chamber. 
}
\label{fig:figure_S3}
\end{minipage}
\end{figure}

\end{document}